# Spin freezing induced giant exchange bias in a doped Hund's metal

S. J. Li[1], D. Zhao[1], J. Li[1], B. L. Kang[1], M. Shan[1], Y. B. Zhou[1], X. Y. Li[1], T. Wu[1,2,3,4,5]* and X. H. Chen[1,2,3,4,5]*

1. Hefei National Research Center for Physical Sciences at the Microscale, University of Science and Technology of China, Hefei, Anhui 230026, China

2. CAS Key Laboratory of Strongly Coupled Quantum Matter Physics, Department of Physics, University of Science and Technology of China, Hefei, Anhui 230026, China

3. CAS Center for Excellence in Superconducting Electronics (CENSE), Shanghai 200050, China

4. Collaborative Innovation Center of Advanced Microstructures, Nanjing University, Nanjing 210093, China

5. Hefei National Laboratory, University of Science and Technology of China, Hefei 230088, China

*Correspondence to: wutao@ustc.edu.cn (T. Wu); chenxh@ustc.edu.cn (X. H. Chen)

**Exchange bias (EB) is a fundamental phenomenon in widespread information technologies. However, a comprehensive understanding of its microscopic origin remains a great challenge. One key issue in the debate is the role of frustration and disorder in the EB mechanism, which motivates the exploration of the EB effect in spin glass (SG) systems. Here, in the SG state of Cr-doped Hund's metal $CsFe_2As_2$, we discover a giant EB effect with a maximum bias field of ~ 2 Tesla, which is almost two orders of magnitude larger than that of traditional alloy SGs. Our results indicate that the giant EB effect should originate from the exchange interactions at the natural boundaries between the tunable ferromagnetic-like (FM) regions around Cr dopants and the SG matrix, via which the FM spins are strongly pinned by the frozen spins in the SG matrix. In addition, the temperature-dependent and cooling-field-dependent EB behaviors could be interpreted well by the SG model with frustrated FM/SG boundaries, which provides an intuitive and explicit understanding of the impact of glassy parameters on the EB effect. All these results suggest that the correlated metals are promising directions for exploring the EB effect in the SG state.**

Exchange bias (EB) is a phenomenon associated with the exchange anisotropy created at magnetic interfaces [1,2] and generally manifests as a shift in the magnetic hysteresis (MH) loop along the field and magnetization axes after field cooling or virgin magnetization processes [3]. Due to its utility in information technologies [4,5], the EB effect has been extensively studied in different systems, including conventional bilayers, fine particles and inhomogeneous bulk materials without well-defined interfaces [6-8], where the uncompensated moments at the pinning boundaries play an important role in determining the bias magnitude. Over decades of studies, different theoretical models have been proposed to interpret the EB phenomenology, such as the basic Stoner-Wolhfarth model, Meiklejohn and Bean (MB) model, random-field model and so on [9-11]. However, a complete understanding of the EB mechanism is still lacking due to complex interfacial magnetism [1,2].

With the continuous emergence of new experimental results, it has gradually been realized that frustration and disorder are essential for generating the EB effect even in typical bilayer systems [10,12,13]. To broaden the understanding of the role played by the glassy order parameters, various EB systems involving the SG phase have been deliberately explored since frustration and disorder are two integral factors in spin-glass systems [14-17]. In addition, other inhomogeneous systems in which the frustrated interfaces between different magnetic phases control the EB behavior have also been investigated [18-20]. The EB effect can even be observed in materials with a single SG phase due to its metastability, but the bias field is generally very small, on the order of 0.01 T [21,22].

In this letter, we observe a giant EB effect in Cr-doped Hund's metal $CsFe_2As_2$, which was recently reported as a new class of spin-glass material with strong electronic correlations [23]. Below the spin-freezing temperature ($T_f$), the spin moments in the ferromagnetic-like (FM) regions introduced by Cr dopants are strongly pinned by the frozen spins in the SG matrix at the frustrated FM/SG boundaries, resulting in the EB effect. At 2 K, the maximum bias field reaches 2 T, belonging to the largest category of reported inhomogeneous EB systems [16,20,24]. Due to frustration and disorder, the bias field and coercive field exhibit nontrivial temperature- and cooling-field-dependent behaviors, which can be interpreted well by the SG model [11,25]. In addition, the giant EB effect shows a distinct doping-content $x$-dependence in Cr-doped $CsFe_2As_2$, where the balance between the volume of the FM regions and the SG matrix tunes the strength of the interfacial pinning effect. These findings provide a new platform for further understanding the role of glassy parameters in the EB mechanism and suggest a possible direction for searching new giant EB materials in correlated metals.

**Evidence for a spin-glass state**

In Fig. 1a, the in-plane magnetic susceptibility $\chi_{ab}$ shows a peak-like behavior and a clear bifurcation between the field cooling (FC) and zero-field cooling (ZFC) curves below the characteristic temperature $T_f$. With increasing magnetic field, $T_f$ shifts toward lower temperatures, as indicated by the arrows. Quantitatively, such field-dependent $T_f$ could be described well by the De Almeida-Thouless (AT) line: $H(T_f) = \Delta J \left[1 - \frac{T_f(H)}{T_f(0)}\right]^{\frac{3}{2}}$, as evidenced by the good linear relationship between $\mu_0 H^{2/3}$ and $T_f$ (Fig. 1b), where $H$ is the applied magnetic field, $T_f(0)$ is the zero-field characteristic temperature and $\Delta J$ is the width of the distribution of exchange interactions [17]. As usually observed in SG systems [15,26], these behaviors support the existence of a spin-glass state in Cr-doped (x = 0.06) $CsFe_2As_2$. The characteristic temperature $T_f$ should be denoted as the spin-freezing temperature, below which the irreversibility between the FC and ZFC curves indicates the presence of an uncompensated frozen moment. In Fig. 1c, another characteristic behavior of the SG state is also detected in Cr-doped (x = 0.06) $CsFe_2As_2$, where the magnetization relaxes with time upon removal of the cooling field below its onset temperature (~ $T_f$), suggesting that the frozen moments are long-term correlated in the spin-freezing state [17,27] (for more details, see Supplementary Section S1). Similar results of magnetization measurements are also detected for Cr-doped (x = 0.10) $CsFe_2As_2$ (see Supplementary Figs. S3b and S3c), suggesting the presence of a SG state consistent with our previous work [23].

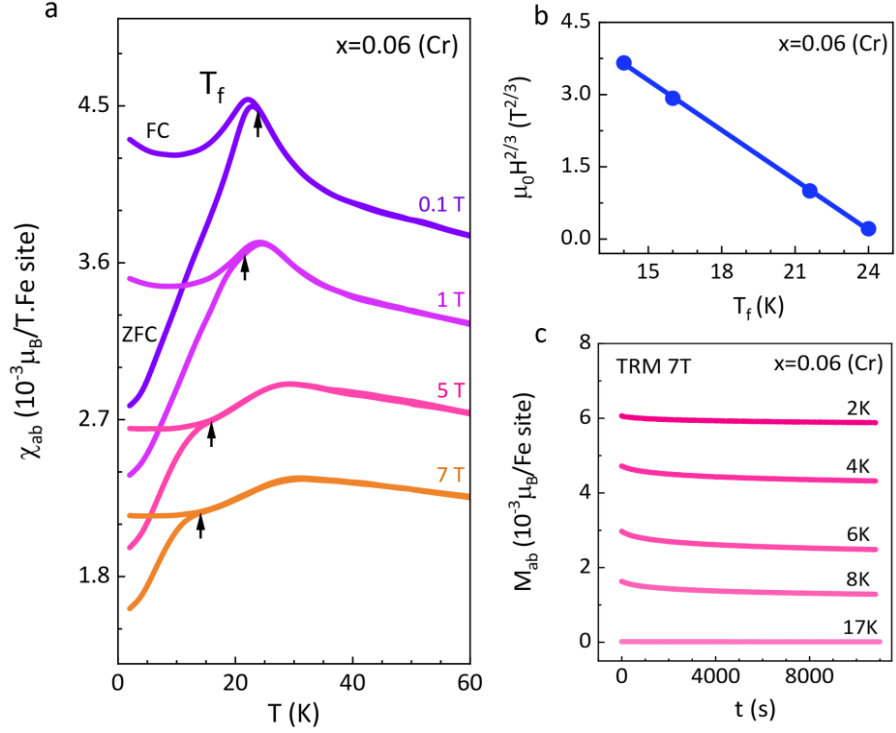

**Fig. 1 | Evidence for a spin-glass state in Cr-doped (x = 0.06) CsFe$_2$As$_2$. a,** Temperature-dependent in-plane magnetic susceptibility $\chi_{ab}$ measured under both field cooling (FC) and zero-field cooling (ZFC) modes with applied fields of 0.1, 1, 5 and 7 T from top to bottom. The black arrows mark the spin-freezing temperature $T_f$, below which the FC and ZFC curves begin to bifurcate (for more details, see Supplementary Fig. S4). For clarity, the MT curves are shifted along the y-axis (for the raw data, see Supplementary Fig. S3a). **b,** Good linear relationship between the physical quantities $(\mu_0 H)^{2/3}$ and $T_f$, confirming the presence of a SG state in Cr-doped (x = 0.06) CsFe$_2$As$_2$. The solid line is the best linear fitting line, from which the obtained $T_f(0)$ (the spin-freezing temperature at $H = 0$) is approximately 24.6 K, which is in agreement with the series of experimental values. **c,** Thermoremanent magnetization (TRM) measurements performed at various temperatures after the sample was cooled from above $T_f$ with a cooling field of 7 T ($H_{FC} = 7$ T). Here, the zero time is defined as the moment when the applied cooling field is removed after waiting for an hour at the target temperature (for full TRM measurement sequences, see Supplementary Section S1). Obviously, the magnetization relaxes with time rather than rapidly decreases to zero after removing the applied field, further confirming the presence of the SG state in Cr-doped (x = 0.06) CsFe$_2$As$_2$.

**Exchange bias effect**

In Cr-doped CsFe$_2$As$_2$, the uncompensated frozen moments could open up a magnetic hysteresis (MH) loop below $T_f$, as shown in Fig. 2. Notably, the loop center shifts away from the origin to the negative field by 2 T at 2 K in Cr-doped (x = 0.06) $CsFe_2As_2$ with a positive cooling field ($H_{FC} = 7$ T) (Fig. 2a), indicating that a strong unidirectional anisotropy occurs in the low-temperature regime. Meanwhile, in Cr-doped (x = 0.10) $CsFe_2As_2$, the loop center shifts to the positive field by 1.4 T with $H_{FC} = -7$ T (Fig. 2d), almost opposite to that with $H_{FC} = 7$ T (Fig. 2c). Moreover, a spontaneous bias has also been observed in Cr-doped (x = 0.06) $CsFe_2As_2$, which refers to the loop shifts without the assistance of cooling fields [3]. In Fig. 2b, the zoomed-in view of the ZFC MH loop clearly shows that such loop shifts are path dependent: in the positive protocol (0 → +7 → −7→ +7 T) (P-type), the loop shifts to the negative field (~− 0.01 T); in the negative protocol (0 → −7 → +7→ −7 T) (N-type), the loop shifts to the positive field (~ 0.01 T). As described in different systems [15,16,24], these observations are common characteristics for an EB effect.

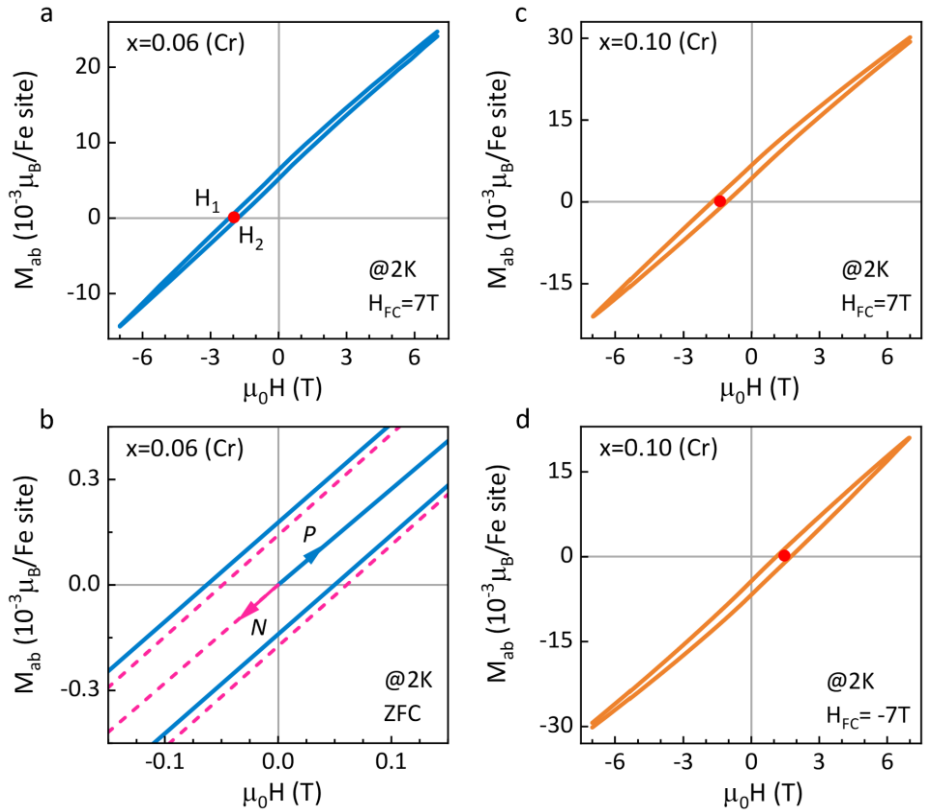

**Fig. 2 | EB effect in Cr-doped $CsFe_2As_2$. a,** Magnetization versus field (MH) curve taken at 2 K in Cr-doped (x = 0.06) $CsFe_2As_2$ after it is cooled from above $T_f$ with a cooling field of 7 T ($H_{FC} = 7$ T). Here, $H_1$ ($H_2$) is the left (right) intercept of the MH loop on the zero-magnetization axis, from which the exchange bias field ($H_{EB} = -\frac{(H_1+H_2)}{2}$) and coercive field ($H_C = \frac{(H_2-H_1)}{2}$) can be obtained. **b,** Zoomed-in view of MH loops taken at 2 K with $H_{FC} = 0$ T in Cr-doped (x = 0.06) $CsFe_2As_2$. Here, the symbol *P* or *N* marks the MH curve in which the sweeping field starts in the positive direction (the solid line) or the negative direction (the

dashed line). The full-field sweeping range is from 7 T to –7 T ($H_{FS} = 7$ T). For clarity, we only present the low-field data. **c, d,** Magnetization versus field (MH) curves taken at 2 K in Cr-doped (x = 0.10) CsFe$_2$As$_2$ with a positive cooling field ($H_{FC} = 7$ T) and negative cooling field ($H_{FC} = -7$ T), respectively. The red solid dot marks the center of each MH loop on the field axis. Clearly, the bias field is symmetric about the origin in the two loops measured with opposite cooling fields.

In EB systems, one another interesting characteristic is known as the training effect (TE), in which the exchange bias field $H_{EB}$ (defined as the average of zero-magnetization intercepts) or coercive field $H_C$ (defined as the half width of the MH loop on the zero-magnetization axis) decreases in repeated magnetization measurements due to the relaxation of interfacial spin configurations [1,28]. As shown in Fig. 3a, $H_{EB}$ gradually decreases with increasing field cycle number $n$ in Cr-doped ($x = 0.06$) CsFe$_2$As$_2$, confirming the presence of TEs. Typically, the data for $n \geq 2$ can be described well by the empirical power law (shown by the gray solid line in Fig. 3a): $H_{EB}(n) - H_{EB}(\infty) \propto \frac{1}{\sqrt{n}}$, where $H_{EB}(\infty)$ is the exchange bias field at the limit of $n \to \infty$ [28]. According to the fitting results, $H_{EB}(\infty)$ is about 1.78 T, which is 10% less than that at $n = 1$, suggesting the stability of the giant EB effect in the training process. In addition, the change in the left intercept ($H_1$) is evidently larger than that in the right intercept ($H_2$) (Figs. 3b-3d) suggests the TE is asymmetric in Cr-doped ($x = 0.06$) CsFe$_2$As$_2$. Similar behaviors have also been detected in Cr-doped ($x = 0.10$) CsFe$_2$As$_2$ from continuous magnetization measurements (see Supplementary Fig. S8). Combined with the cooling-field- and path-dependent shifts, such asymmetric TEs further confirm that the loop shifts should be more attributed to an intrinsic EB effect in Cr-doped CsFe$_2$As$_2$, not the non-saturated minor loops, since they are independent of the magnetic history as reported in ferromagnetic materials [15,16,29]. Even if there is, the minor loop effect would make only a small contribution to the loop shifts, which will has no substantial impact on the following discussions.

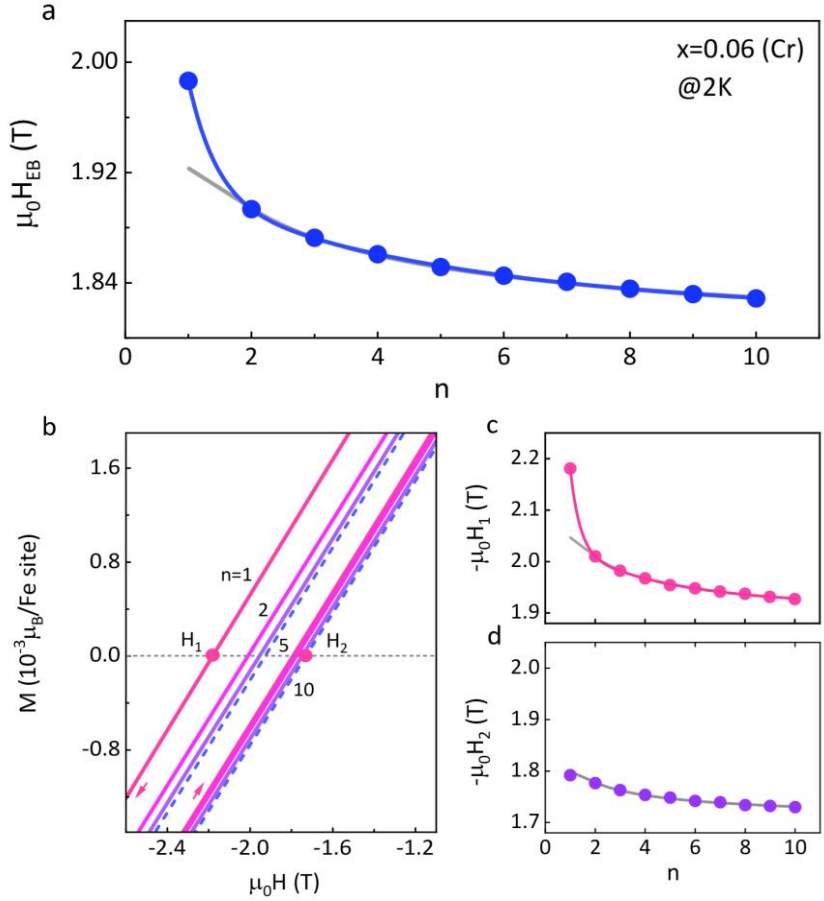

**Fig. 3 | Training effect in Cr-doped (x = 0.06) CsFe$_2$As$_2$. a,** Evolution of $H_{EB}$ with $n$, where $n$ is the field cycle number in continuous magnetization measurements. With increasing $n$, $H_{EB}$ gradually decreases, indicating the presence of TEs. **b,** Zoomed-in view of continuous MH loops taken at 2 K after the sample is cooled from above $T_f$ with $H_{FC} = 7$ T, starting in the positive direction, as indicated by the arrows. The full-field sweeping range is from 7 to – 7 T ($H_{FS} = 7$ T). For clarity, we only show several representative MH loops with $n = 1, 2, 5$, and 10 from left to right. $H_1$ ($H_2$) is the left (right) intercept of the MH loop on the zero-magnetization axis. **c, d,** Evolutions of $-H_1$ and $-H_2$ with $n$, respectively. For comparison, the value range of the vertical coordinate in panel c is the same as that in panel d. The gray solid line in panels a, c or d shows the best fit with the power law to the $n$-dependent $H_{EB}$, $-H_1$ or $-H_2$ for $n \geq 2$, which has been extrapolated to $n = 1$. Clearly, the power law cannot be used to describe the reduction in $H_{EB}$ between $n = 1$ and $n = 2$. The blue (pink) solid line is the best fitting curve to the $n$-dependent $H_{EB}$ ($-H_1$) for $n \geq 1$ according to the frustrated model, as mentioned in the text. Here, $-H_2$ could not be fitted well by the frustrated model due to the small change with $n$.

**Spin-glass model with natural FM/SG boundaries**

From the further systematic magnetization measurements performed on Cr-doped (x = 0.06) and (x = 0.10) CsFe$_2$As$_2$ (see Supplementary Figs. S5 and S7), the obtained parameters $H_{EB}$ and $H_C$ show distinct temperature-dependent behaviors. As shown in Fig. 4a, $H_{EB}$ starts to be nonzero and increases below $T_f$ ($T_f$ ~ 14 K, $H_{FC}$ = 7 T). That is, the blocking temperature $T_B$ is very close to $T_f$ in Cr-doped CsFe$_2$As$_2$, above which the EB effect vanishes, as is previously defined [1,2]. Moreover, the onset temperature of nonzero $H_C$ is slightly greater than $T_f$ ($T_B$) (Fig. 4b), as is usually observed in other EB systems [1,11]. In sharp contrast, no signals for the EB effect could be detected in the pristine and Co-doped CsFe$_2$As$_2$, where the SG state is absent and instead, the emergent Fermi-liquid state is present at low temperatures [23] (for more details, see Supplementary Section S6). These observations powerfully indicate that the giant EB effect is tightly intertwined with the spin-freezing state in Cr-doped CsFe$_2$As$_2$. As usually reported in other iron-based materials [30,31], the FM correlation could be enhanced with doping of Cr atoms, which has been evidenced in Cr-doped CsFe$_2$As$_2$ from the apparent increase of Curie temperature obtained by the Curie-Weiss fitting for high-temperature magnetic susceptibilities [23] (see Supplementary Fig. S11 of Ref. 23). Hereto, it is reasonable to suppose that the exchange interactions at the pinning boundaries between the FM regions induced by Cr dopants and the SG matrix are accountable for the observed EB effect in Cr-doped CsFe$_2$As$_2$. Moreover, $H_{EB}$ and $H_C$ exhibits an exponential decay with increasing temperature (Figs. 4a and 4b), further supporting the above picture since they are typical behaviors for EB systems with frustrated FM/SG pinning boundaries [32,33].

However, not all the interfacial spins in the spin-freezing state contribute to the EB effect due to its frustrating and disordered properties. One part, "frozen spins", results in a biasing field, remaining mainly intact with a large anisotropy on the reversal of FM spins; another part, named "rotatable spins", gives rise to an additional coercivity, following the rotation of FM spins with a smaller anisotropy, as proposed in the SG model (one modified MB model) [11,25]. A supportive evidence for this scenario is that the $n$-dependent $H_{EB}$ could be fit well by the following frustrated model proposed by Mishra *et al.* for $n \geq 1$ (shown by the blue solid line in Fig. 3a): $H_{EB}(n) - H_{EB}(\infty) = A_f \, exp\left(-\frac{n}{P_f}\right) + A_r exp\left(-\frac{n}{P_r}\right)$ [34]. Here, $A_f$ and $P_f$ are the parameters related to the frozen spins, $A_r$ and $P_r$ are the parameters related to the rotatable spins, and the *A* parameters have unit of Oersted, while the *P* parameters are dimensionless but resemble a relaxation time. It is not possible to describe the $n$-dependent $H_{EB}$ by only one exponential. According to the fitting results, $H_{EB}(\infty)$ is about 1.82 T, close to that obtained from the power-law fitting; the other parameters are $A_f = 0.10 \pm 0.003$ T, $P_f = 4.59 \pm 0.414$, $A_r = 0.81 \pm 0.112$ T and $P_r = 0.45 \pm 0.034$. It is clear that the frozen spins appear to relax about 10 times slower than the rotatable spins. In such a frustrated model, changes in both frozen and rotatable components could modify the spin configurations at the pinning boundaries. With each cycle, the interfacial spin disorder increases, causing a decrease in $H_{EB}$ or $H_C$ in the training process [11,34].

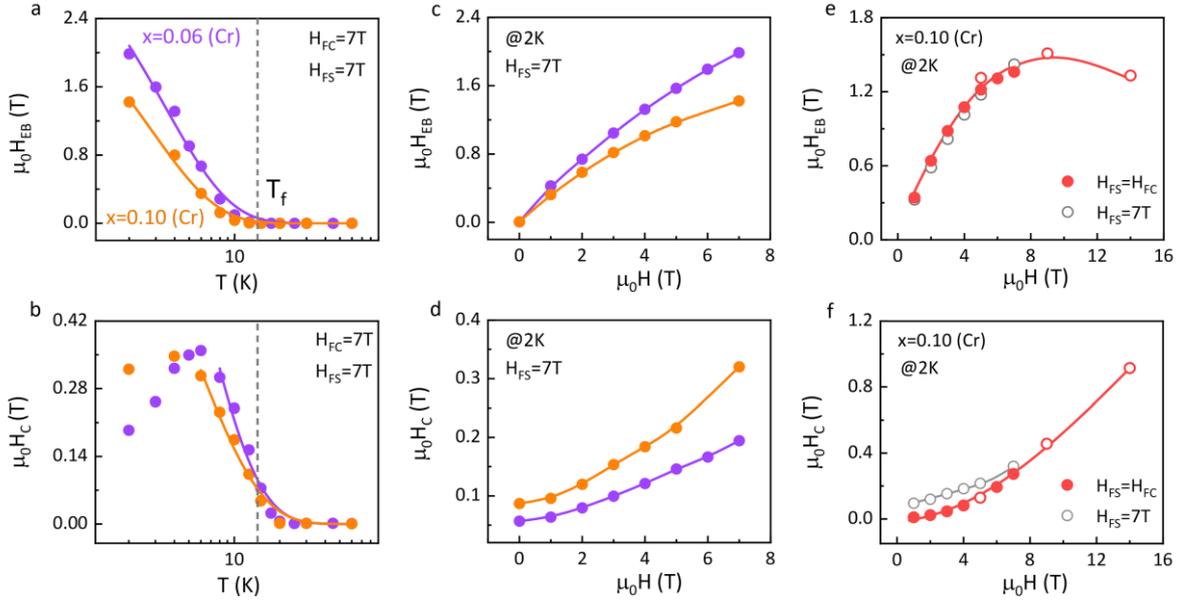

**Fig. 4 | Evolutions of $H_{EB}$ and $H_C$ in Cr-doped CsFe$_2$As$_2$. a, b,** Temperature-dependent $H_{EB}$ and $H_C$ in Cr-doped CsFe$_2$As$_2$ (x = 0.06 (purple dots) and x = 0.10 (orange dots)) extracted from hysteresis loops measured with $H_{FC}$ =7 T and $H_{FS}$ =7 T (see Supplementary Figs. S5 and S7). In panels a and b, the gray dashed line marks the spin-freezing temperature $T_f$, below which $H_{EB}$ starts to be nonzero and increase. The solid lines are the best fitting curves to the temperature-dependent $H_{EB}$ and $H_C$ (above peak temperature) according to the exponential decay formula: $H_{EB}(H_C) = H_0 \exp(-\frac{T}{T_0})$, where $H_0$ is the extrapolation of $H_{EB}(H_C)$ at 0 K and $T_0$ is a constant [32,33]. **c, d,** Cooling-field-dependent $H_{EB}$ and $H_C$ in Cr-doped CsFe$_2$As$_2$ (x = 0.06 (purple dots) and x = 0.10 (orange dots)) extracted from hysteresis loops measured at 2 K with $H_{FS}$ =7 T. In the range of $0 \leq H_{FC} \leq 7$ T, both $H_{EB}$ and $H_C$ increase with increasing $H_{FC}$. **e, f,** Cooling-field-dependent $H_{EB}$ and $H_C$ in Cr-doped (x = 0.10) CsFe$_2$As$_2$ (red dots), extracted from hysteresis loops measured at 2 K with $H_{FS} = H_{FC}$ and $0 < H_{FC} \leq 14$ T. The gray dots in panles e and f are the same as the data in panels c and d (orange dots), respectively. The hollow red signals are measured on the PPMS-VSM (14 T, Quantum Design), while the solid red signals are measured on the SQUID-VSM (7 T, Quantum Design) (for MH loops, see

Supplementary Fig. S10). Clearly, these two sets of data could be well interlinked in Cr-doped (x = 0.10) CsFe$_2$As$_2$ at 2 K. The solid lines in panels c-f can guide the eyes.

**Temperature dependence of the EB effect**

In contrast to conventional FM/AFM bilayers with well-defined interfaces, the number of FM/SG boundaries is changeable in such frustrated systems, which is crucial for tuning the bias field. In addition, the ratio between frozen and rotatable spins, magnetic anisotropy of different constituents (FM region: $K_F$, SG matrix: $K_{SG}$) and doping content of Cr atoms ($x$) also have significant effects on the behaviors of $H_{EB}$ or $H_C$ [1,3,11]. As shown in Fig. 4a, $H_{EB}$ increases below $T_f$, mainly due to the increase in $K_{SG}$ with decreasing temperature. Moreover, an increase in $K_F$ causes an increase in $H_C$ upon cooling below the onset temperature (Fig. 4b). Keep in mind that the additional coercivity coming from the rotatable spins at the pinning boundaries also contributes to the total coercive field $H_C$. Below $T_f$, a remarkable increase in $H_{EB}$ means that an increasing number of the rotatable spins are converted to the frozen spins, enhancing the pinning force on the FM spins while reducing the additional coercivity. The combined effect of the opposite temperature-dependent coercivity contributions from the FM spins and rotatable spins leads to the peak behavior of $H_C$ (Fig. 4b).

By further comparison, the values of $H_{EB}$ and $H_C$ are apparently different for the two Cr-doped CsFe$_2$As$_2$. With an increase in $x$ from 0.06 to 0.10, the number of FM regions introduced by Cr dopants increases, while the volume fraction of the SG matrix clearly decreases. In this case, more spins of the SG phase can be "dragged" by the FM spins via exchange interactions, meaning that the interfacial pinning effect could be weakened [3]. Therefore, the exchange bias field $H_{EB}$ is smaller in Cr-doped ($x = 0.10$) CsFe$_2$As$_2$ (Fig. 4a). In addition, the magnetization of FM regions ($M_F$) also increases with increasing $x$, as seen clearly by comparing the temperature-dependent $\chi_{ab}$ shown in Figs. S3a and S3b (Supplementary Information). Due to the increase in magnetization, $H_C$ of Cr-doped ($x = 0.10$) CsFe$_2$As$_2$ is also smaller than that of Cr-doped ($x = 0.06$) CsFe$_2$As$_2$ (Fig. 4b) despite the larger number of FM regions according to this expression: $H_C \propto 2K_F/M_F$ [11]. However, this situation reverses below the peak temperature of $H_C$ because fewer rotatable spins convert to the frozen spins in Cr-doped ($x = 0.10$) CsFe$_2$As$_2$, as evidenced by its smaller value of $H_{EB}$ at very low temperatures (Fig. 4a).

**Influence of the cooling field on the EB effect**

Beyond that, the volume fraction of FM regions and SG matrix could be tuned by the cooling field ($H_{FC}$) via reconfiguring the spin moments, generating a prominent impact on EB behaviors [3,35]. With increasing $H_{FC}$, the FM spins and interfacial spins in the spin-freezing state are aligned more along the direction of the cooling field, reducing the averaging effect of exchange anisotropy due to randomness and enhancing the interfacial bias effect. In addition, the size and number of FM regions increase at the expense of the SG matrix driven by the greater Zeeman energy [1], resulting in the growth of $H_C$ with $H_{FC}$ ($0 \leq H_{FC} \leq 7$ T) (Fig. 4d). Simultaneously, more FM/SG pinning boundaries are formed in Cr-doped CsFe$_2$As$_2$, resulting in an increase in $H_{EB}$ with $H_{FC}$ ($0 \leq H_{FC} \leq 7$ T) (Fig. 4c). When the cooling field is high enough, the localized FM regions coalesce into larger coherent ones by destroying the spin-freezing state once the interaction between FM regions reaches the critical value [36]. At this moment, the proportion of the SG matrix and FM regions is sharply reduced, meaning that the few remaining spins in the spin-freezing state cannot readily pin the FM spins. In other words, $H_{EB}$ begins to decrease beyond the critical cooling field, reaching approximately 9 T in Cr-doped ($x = 0.10$) CsFe$_2$As$_2$ (Fig. 4e). However, the coercive field $H_C$ continues to increase until the cooling field increases to 14 T (Fig. 4f). If the cooling field increases further, it is possible that such EB effect would even disappear in Cr-doped CsFe$_2$As$_2$ when the spin-freezing state is completely destroyed under the action of sufficiently large Zeeman energy. Here, $H_{EB}$ remains at approximately 1.3 T with $H_{FC} = 14$ T, indicating the robustness of the EB effect under a relatively high field in Cr-doped CsFe$_2$As$_2$.

**Discussion**

Similar EB phenomena have also been reported in other systems with single phase of SG state or spin-glass interface, which is dominant in modulating the magnitude and behavior of the EB effect [21,26,32,33,36-39]. Among them, most exchange bias fields are generally on the order of 0.01 to 0.1 T, as summarized in Fig. 5 (the purple dots), one or two orders of magnitude smaller than the giant value of 2 T in Cr-doped ($x = 0.06$) CsFe$_2$As$_2$. As it is known, there is no theoretical model that could explain all the EB phenomena well due to the diversity of influencing factors at present. However, the limit value of different models under the ideal condition can be expressed as $H_{EB} \propto \frac{J_{eb}}{\mu_0 t_F M_F}$, similar to that given by the ideal MB model based on typical ferromagnet/antiferromagnet bilayers [9,11]. Here, $\mu_0$ is the vacuum permeability, $J_{eb}$ is the interfacial exchange coupling parameter, $M_F$ is the saturation magnetization of FM layer and $t_F$ is the thickness of FM layer. This clearly shows that the ratio $J_{eb}/t_F M_F$ is an important factor in determining the magnitude of $H_{EB}$, albeit other different parameters might be involved in the specific expressions of different models [11]. In spin-glass systems, $J_{eb}$ should depend on the exchange interaction constant ($J_{SG}$) between spin moments; $M_F$ and $t_F$ should stand for the magnetization and size of FM regions or clusters respectively.

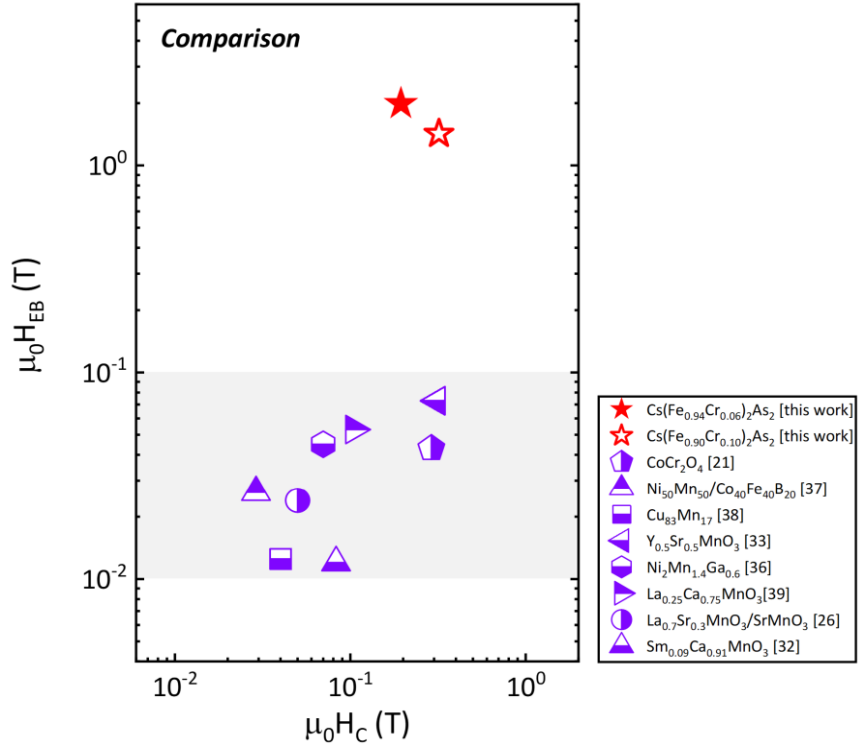

**Fig. 5 | Comparison of $H_{EB}$ and $H_C$ among various exchange-biased systems with a single phase of SG state or spin-glass interface, which is dominant in modulating the magnitude and behavior of the EB effect.** It is evident that the maximum bias field of 2 T in our Cr-doped ($x = 0.06$) CsFe$_2$As$_2$ (the solid red star) belongs to the largest category of the reported frustrated EB systems.

As reported in our earlier work [23], the SG state is consistent with the canonical SG in Cr-doped CsFe$_2$As$_2$ as evidenced from the frequency-dependent ac magnetic susceptibility measurements (the characteristic relaxation time of spin dynamics $\tau_0 \sim 10^{-12}$ s and the parameter $\delta$ is on the order of 0.001) (for more details, see Supplementary Section S7). In canonical SG, the dynamical relaxation is dominated by the single-ion flipping process and magnetic spin textures spread like a collective excitation through the whole sample due to the strong spin correlation [17,40]. Whereas, most of SG systems are reported to be cluster SGs ($\tau_0 \gtrsim 10^{-11}$ s, $\delta \sim 0.01$) [21,36,38,39], in which the dynamical relaxation arises from the motion of domain walls or clusters and the correlation length is restricted to the size of domains or clusters due to the moderate correlation. Such contrast indicates that $J_{SG}$ should be larger in the canonical SG matrix of Cr-doped CsFe$_2$As$_2$ with the strong exchange interactions between spin moments. At the meantime, $t_F$ should be obviously smaller in the strong-correlated Cr-doped CsFe$_2$As$_2$ since $t_F$ is positively related to $\tau_0$ in general [41]. Here, we call these FM coupled assemblages of spins regions rather than clusters due to their small sizes. Whether there are microscopically measurable clusters in Cr-doped CsFe$_2$As$_2$ needs further high-resolution magnetic force microscopy measurements to verify. Moreover, the need of very large field to saturate the magnetization (seen from the almost linearly increasing trend of magnetization with field up to 7 T or even 14 T) suggests $M_F$ is weak in Cr-doped CsFe$_2$As$_2$ because the smaller $M_F$, the smaller the Zeeman energy felt by spins [42]. Hereto, a great ratio $J_{eb}/t_F M_F$ can be acquired in Cr-doped CsFe$_2$As$_2$ in light of the larger $J_{SG}$ and smaller value of $t_F M_F$, which could contribute to the giant EB effect in Cr-doped CsFe$_2$As$_2$.

In previous reports, a few systems are also consistent with canonical SGs, but the maximum $H_{EB}$ could only reach the order of KOe [26,33]. It may be because the object of survey is nanoparticle or polycrystal, whose average anisotropic energy could be smaller due to the randomness [33]. Another is possibly on account of the small interfacial volume ratio since the SG state is only confined to a thin-film interface [26], whereas, the FM/SG pinning boundaries are throughout the whole volume of sample in Cr-doped CsFe$_2$As$_2$. Moreover, in these oxides or alloys, the occurrence of SG state mainly originates from the long-range Ruderman-Kittel-Kasuya-Yosida (RKKY) interactions [17,40]. In our study, the Cr-doped CsFe$_2$As$_2$ belongs to the category of Hund's metal, which exhibits the remarkable orbital differentiation on the electronic correlations under the prominent influence of Hund's coupling [43,44]. Due to the orbital selectivity, unusual magnetic frustration might be introduced by the competition among different magnetic excitations in the different $3d$ orbitals as reported in FeSe [45], which would suppress the formation of long-range magnetic orders and promote the appearance of SG state in Cr-doped CsFe$_2$As$_2$. Such probable difference in the origin of SG state may be also responsible for the giant EB effect in the correlated Hund's metal, which requires the further theoretical development for a elaborate interpretation.

In conclusion, we reveal a giant EB effect in Cr-doped CsFe$_2$As$_2$, which can be tuned by changing the number of pinning boundaries, volume fraction of the FM regions and the SG matrix or the ratio of rotatable and frozen spins at the frustrated FM/SG pinning boundaries via the cooling field or doping content. Different from other reported giant EB materials [16,20,24], the giant EB effect is observed in Cr-doped Hund's metal CsFe$_2$As$_2$, which presents a single phase of SG state without other long-range magnetic orders as verified from the characterizations of different measurements [23]. Moreover, it is the first time that such a giant EB effect of 2 T is detected in the exchange-biased system with a pure SG state, which might result from the strong spin and electronic correlations in the canonical SG of Cr-doped Hund's metal CsFe$_2$As$_2$. These findings could offer a fresh opportunity to better comprehend the EB phenomena in the correlated SG system, meanwhile, point out a possible direction to find new stable giant EB materials in such correlated metals for potential applications in spintronics.

## Methods

**Sample preparation and characterization.** High-quality Cr-doped and Co-doped CsFe$_2$As$_2$ single crystals were synthesized by the self-flux method, as reported in Ref. 23. In this paper, the actual doping contents of Cr (Co)-doped CsFe$_2$As$_2$ are x = 0.06 (x = 0.07) and x = 0.10 (x = 0.11) with nominal doping contents of x = 0.10 and x = 0.15, respectively. Detailed information on the X-ray diffraction (XRD) patterns and results of the energy dispersive X-ray spectroscopy (EDS) measurements performed on the different doped samples were reported in our previous work [23].

**Magnetization measurements.** Magnetization measurements were measured on a commercial SQUID-VSM (7 T, Quantum Design) ($0 \leq H \leq 7$ T) and PPMS-VSM (14 T, Quantum Design) ($5 \leq H \leq 14$ T). Due to the weak magnetic signal of Cr (Co)-doped CsFe$_2$As$_2$, a quartz paddle sample holder was used to minimize the magnetic background. The doped sample was glued to the holder in a horizontal orientation with a small amount of GE varnish and then removed from the glove box with a heat seal bag. Since these samples are very sensitive to air and moisture, the entire process of sample preparation was carried out in a glove box ($H_2O < 0.1$, $O_2 < 0.1\ ppm$). Moreover, the residual cosolvent on the surface had a certain effect on the magnetic susceptibility behavior of the doped samples, which needed to be quickly wiped off with a small amount of alcohol in the glove box.

## Additional Information

Supplementary Information is available online or from the author.

## Acknowledgments

This work is supported by the National Key R&D Program of the MOST of China (Grant No. 2022YFA1602601), the National Natural Science Foundation of China (Grants No. 12034004, 12161160316, 12325403), the Chinese Academy of Sciences under contract No. JZHKYPT-2021-08, the CAS Project for Young Scientists in Basic Research (Grant No. YBR-048), and the Innovation Program for Quantum Science and Technology (Grant No. 2021ZD0302800).